\documentclass{jpsj-suppl}
\usepackage{txfonts} 

\title{Nuclear Structure Studies of Neutron-Rich Nuclei: Astrophysical Implications}

\author{Olivier \textsc{Sorlin}$^{1}$}

\inst{$^{1}$ Grand Acc\'el\'erateur National d'Ions Lourds (GANIL),
CEA/DSM - CNRS/IN2P3, B.\ P.\ 55027, F-14076 Caen Cedex 5, France}

\email{sorlin@ganil.fr}

\recdate{August, 20 (2016)}

\abst{It is proposed here to investigate three major properties of the nuclear force that influence the amplitude of shell gaps, the nuclear binding energies as well as the nuclear $\beta$-decay properties far from stability, that are all key ingredients for modeling  the r-process nucleosynthesis. These properties are derived from experiments performed in different facilities worldwide, using several various state-of-the-art experimental techniques including transfer and  knockout reactions.  Expected consequences on the r process nucleosynthesis as well as on the stability of super heavy elements are discussed.}

\kword{spin-orbit force, drip-line, r-process, transfer and knock-out reactions}

\begin{document}
\maketitle

\section{Introduction}

A suitable description of  the r-process nucleosynthesis requires in particular nuclear structure information from about thousands of nuclei close-to or at the drip line, that are, despite the tremendous progresses achieved in producing rare ions in radioactive ion beam facilities, hardly or not accessible experimentally. This results in large uncertainties in predicting  r-process abundances, that add to those related to the identification of the r-process sites and to their hydrodynamical modeling. Two complementary approaches exist to reach this nuclear structure endeavor. 

\noindent The first consists in measuring properties relevant to the r-process, such as atomic masses, lifetimes, neutron-delayed emission probabilities, and level schemes, for nuclei potentially  paving the r-process path or as close as possible to it. The recent measurement of more than 100 lifetimes along the r-process path \cite{Lorusso} has lead to a significant improvement on the previous deficits in fitting the r-process curve close to the r-process peak $A\simeq$130. These experimental results can then be compared to various theoretical predictions, and from this comparison, those that closely reproduce the experimental data can be selected as being probably the most relevant to be used further away from stability. 

\noindent The second consists in finding suitable nuclei, not necessarily involved in the r-process, in which the relevant  properties of the nuclear force can be tested and constrained, in order to be applied to yet unexplored regions of the chart of nuclides.  In this review, it is proposed in {\it Sect.\ref{one}} to propose a way to constraint the density and isospin dependence of the spin-orbit force toward the neutron drip line and in the region of super-heavy elements using the bubble nucleus $^{34}$Si, in {\it Sect.\ref{two}} to reach a better shell evolution predictability in the region of the r-process path based on the hierarchy of nuclear forces derived from the study of nuclei with $A\leq$ 80, in {\it Sect.\ref{three}} to evaluate the change in effective proton-neutron interaction  at the drip line (as compared to stable nuclei), through the study of $^{26}$F. 

\section {Density and isospin dependence of the nuclear spin-orbit force} \label{one}

The spin orbit (SO) force was first introduced in 1949 by M.~Goeppert-Mayer and O.~Haxel et al.~\cite{MGMHAXEL} as a result of a strong  $\bf{\ell} \cdot \bf{s}$ coupling, that is  attractive for nucleons having their orbital angular momentum aligned with respect to their spin ($j _\uparrow=\ell+s$) and repulsive in case of anti-alignment ($j _\downarrow=\ell-s$). Large gaps are created between the $j_\uparrow$ and $j_\downarrow$ orbits at nucleon numbers 6, 14, 28, 50, 82 and 126 for $\ell$=1-6,  which could not be explained otherwise. The shell gaps 50, 82 and 126 are strongly connected to the s and r-process abundance peaks and  their reduction far from the valley of stability would influence the position, height and shape of the r-process peaks (see e.g. ~\cite{Pfeiffer}).  

\noindent Relativistic Mean Field (RMF) models later introduced a more fundamental description of the nuclear spin-orbit interaction~\cite{PRing} derived from Dirac equation, whose non-relativistic derivation writes:
\begin{equation}\label{SO}
  V_\tau^{\ell s} (r) = - (W_1 \partial_r \rho_\tau (r) + W_2 \partial_r \rho_{\tau\neq \tau '} (r)) \vec{\ell} \cdot \vec{s}
\end{equation}

In this expression, the  SO interaction contains three important components: the $\vec{\ell} \cdot \vec{s}$ term already introduced earlier to account for shell gaps,  the derivative of the nuclear density $\rho(r)$ for like or unlike ($\tau\neq \tau '$) particles, and the isospin dependence through the $W_{1}/W_{2}$ ratio that depends on the $\sigma, \omega, \rho$ meson coupling constants. The density and isospin dependences of the SO interaction are subject to large uncertainties between relativistic and non-relativistic models with $W_1/W_2$ ratio ranging from 1  to 2 (see e.g. the discussions in \cite{Sharma95,Lala98,Holt11,Bender99}). In particular, there is moderate isospin dependence of the SO interaction in the RMF approach ($W_1/W_2 \simeq$ 1), meaning for instance that a proton depletion will affect in the same manner the proton and neutron SO force.  On the other hand, non-relativistic Hartree-Fock approaches have a large isospin dependence ($W_1/W_2$=2) that would for instance reduce by a factor of two  the effect of a {\it proton} depletion on the {\it neutron} SO force.

\begin{figure} [h]
\begin{center}
\includegraphics[width=12cm] {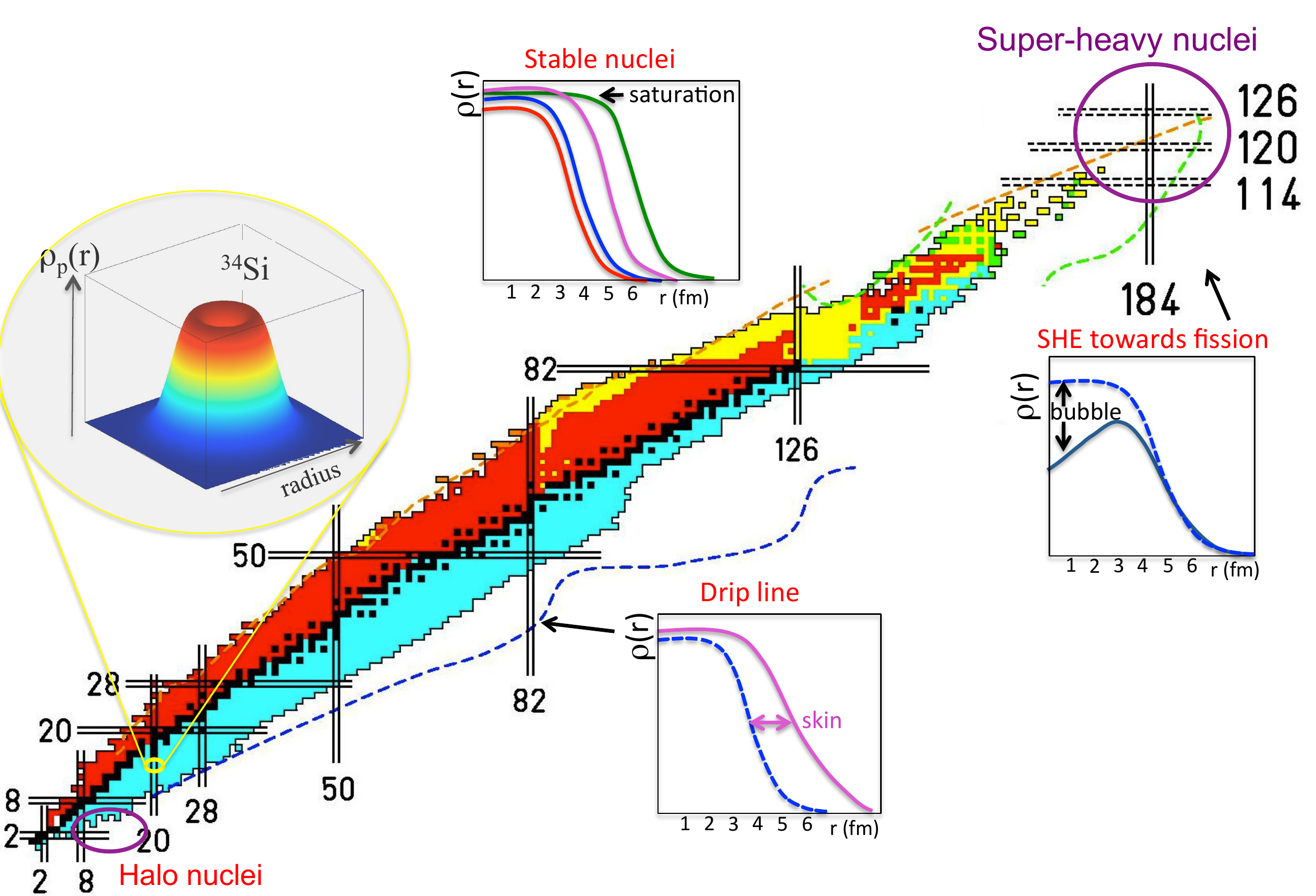}
\end{center}
\caption {Chart of nuclides showing the density distributions determined for stable, expected towards super-heavy elements (SHE) and at the neutron-drip-line. The newly determined proton density distribution of $^{34}$Si, obtained from one-proton knockout reaction, display a central depletion, is visualized here using Relativistic Hartree-Fock Bogoliubov (RHFB) calculations with PKO2 energy density functional \cite{ebr11}, which predict very similar proton and neutron occupancies to those deduced experimentally. 
} \label{density} 
\end{figure}

These components of the SO force could not be constrained experimentally so far, as most of the nuclei studied in the valley of stability similarly exhibit constant proton and neutron density profiles in their interior, and a sharp drop of the same slope at their surface (see Fig. \ref{density}). Significant changes in density profiles are however expected at the drip-line and in the region of super-heavy nuclei. Drip-line nuclei are expected to display increased surface diffuseness (see Fig. \ref{density}) making the derivative of the nuclear density along the surface weaker and thus the SO splitting smaller. This increased diffuseness would cause a reduction of the neutron SO shell-gaps far from the valley of stability, where the r-process occurs, the intensity of which is strongly model-dependent~\cite{Sharma95}. Some of the super-heavy elements (SHE) exhibit a central proton depletion because of the large Coulomb repulsion between protons. In addition, according to RMF calculations, nuclei in the vicinity of $^{292}120_{172}$ are expected to exhibit large proton \emph{and} neutron central density depletions\cite{Bender99, Bender01}, despite being reduced when correlations are taken into account \cite{Afa}. Central depletions adds  a component of the SO interaction in the center of the nucleus of opposite sign to the one at the surface that globally reduced the amplitude of the SO splitting. The amplitude of this reduction in this region of SHE, and herewith the existence of more or less pronounced stabilizing gaps originating from this force, strongly depends on the unconstrained $W_1/W_2$ ratio. Moreover, the mass distribution arising from the fission of heavy or super-heavy elements, essential for modeling fission cycling in neutron-star mergers, also likely depend on the density and isospin dependence of the SO force and its stabilizing effect at the time nuclei undergo scission. 

Though not directly involved in the r-process nucleosynthesis, an ideal candidate to study these poorly constrained properties of the SO force is the $^{34}$Si nucleus.  Being radioactive,  $^{34}$Si was produced in a fragmentation reaction at the NSCL/MSU facility and selected in flight using the A1900 spectrometer. Proton knockout reaction was used in inverse kinematics to determine, from the measured $(-1p$) cross section, that the proton $2s$ occupancy in $^{34}$Si is only 10\% of that of the stable $^{36}$S nucleus \cite{Mutschler}. Being peaked at the center of the nucleus, this lack of $2s$ protons induces a significant central depletion, as shown in the insert of Fig.\ref{density}. On the other hand, neutron density distributions of the two nuclei are similar, without central depletion.  It follows that, with a proton but no neutron central depletion,  $^{34}$Si is especially suited  to study density and isospin dependence of the SO force.

The central proton density depletion in $^{34}$Si drives an additional (interior) component of the SO force, with a 
sign opposite to that at the surface. Therefore, low-$\ell$ nucleons, that can probe the interior of the proton bubble, should encounter a 
weaker overall SO interaction (that  scales with the $W_{1}/W_{2}$ ratio), and display a significantly reduced SO splitting. The $^{34}$Si $(d,p)^{35}$Si reaction was used in inverse kinematics at the GANIL facility to show that the neutron $2p_{3/2} - 2p_{1/2}$ splitting in $^{35}$Si is reduced by a factor of about two \cite{Burg14}, as compared to the neighboring $N=21$ isotones of $^{41}$Ca and $^{37}$S. Such a sudden change, connected to that in central density,  demonstrates for the first time the density-dependent of the SO force. This result  will be used in the future to contraint the strength of SO force in a unique manner, and to explore consequences in the r-process nucleosynthesis and in the modeling of SHE.

\section {Shell evolution and the hierarchy of nuclear forces}  \label{two}

The nuclear force can be casted into central, spin-orbit and tensor parts. When moving along the chart of nuclides, protons and neutrons occupy orbits with different combinations of quantum numbers $(n,\ell,j)$, with $n$ the number of nodes of the radial wave function, $\ell$ the orbital momentum and $j$ the total spin value. It follows that in a given chain of isotones, various components of the proton-neutron nuclear force are explored, inducing either no change in shell structure, or on the contrary a significant reduction of shell gaps and even a shell re-ordering. 

\begin{figure} [h]
\begin{center}
\includegraphics[width=11cm] {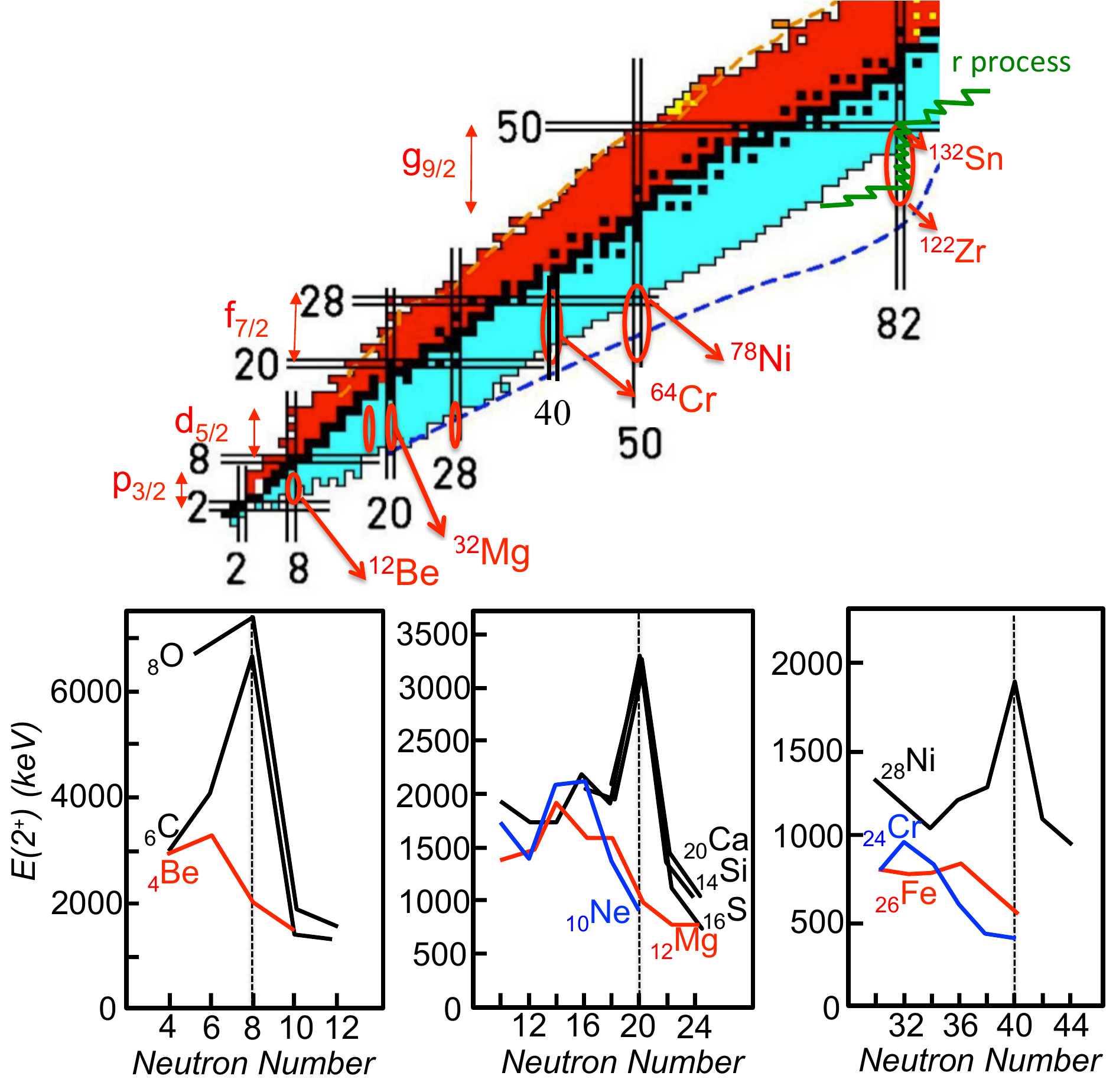}
\end{center}
\caption {{\bf Top}: The regions of the chart of nuclides circled in red correspond to protons having their orbital momentum and spin aligned ($j _\uparrow$). The filling of these orbitals (e.g. $d_{5/2}$, $f_{7/2}$ or $g_{9/2}$) coincide with major changes in the structure of nuclides (in particular their 2$^+_1$ energies) shown in the bottom part of the figure. {\bf Bottom}: The rise of 2$^+_1$ energies at the neutron shell closures $N$= 8, 20 and 40 (black lines)  no longer exist far from the valley of stability (red and blue lines), where a steep decrease of 2$^+_1$ energies is found instead. This apparent breaking of magicity far from stability represents one of the major change of paradigm in nuclear physics that should also apply in regions where the r-process nucleosynthesis takes place.} \label{hierarchy} 
\end{figure}

Shell disappearance or weakening could be viewed at first glance in the evolution of the first excited states (2$^+$) of even-even nuclei as a function of neutron numbers $N$, as shown in the bottom part of Fig. \ref{hierarchy}.  It took about 30 years to establish these curves, from experiments carried out worldwide at radioactive ion beam facilities (see e.g. Refs\cite{Hannawald,Sorlin,Crawford}) at N=40 . It can be seen that, in the valley of stability, all isotopic chains (black curves) display a sudden increase in 2$^+$ energy at the shell closures $N$=8, 20 and 40. The effect is remarkably preserved at $N$=20 from $Z$=20 down to $Z$=14. This observation led to the earlier paradigm that magic nuclei were immutable. On the other hand, far from stability, as seen with the blue and red curves, there is no more sign of increase in 2$^+$ energy, as if all magic numbers no longer exist far from stability. It has been proposed in Ref. \cite{sorlin-h} that these shell gaps disappear in the same manner in all regions, owing in particular to a common hierarchy of forces that is present far from stability but no longer in the valley of stability.  In particular, shell effects increase in an isotonic chain when the proton orbits having their angular momentum aligned with their spin value $j _\uparrow$ are filled, e.g. $f_{7/2}$. They are preserved afterwards. Important is to notice that the same forces are expected to be at play in  particular in all regions circled in yellow in the chart of nuclei of Fig.\ref{hierarchy}. Of particular importance in astrophysics is the region below $^{132}$Sn, where the $A \simeq$130 r-process abundance peak is built. Following this hierarchy, a weakening of the $N$=82 is also expected there. However, as containing a large number of nucleons, the net effect of removing few of the 10 protons occupying the $g_{9/2}$ orbit from $^{132}$Sn will not be as strong as the removal of  for instance two protons from $^{14}$C to $^{12}$Be, where an abrupt change of shell structure is revealed from the sudden drop in 2$^+$ energy at $N$=8 (see Fig. \ref{hierarchy}). The reduction of shell gaps and the reordering of orbits should arise from this hierarchy, impacting for instance the location of the (n,$\gamma$)- ($\gamma$,n) equilibrium in the r-process, as well as $\beta$-decay lifetimes and neutron-delayed emission probabilities P$_n$. Many models and parameterizations are used nowadays to calculate r-process abundances. Some of them may have the suitable ingredients included that lead to this hierarchy, others not, leading to significant differences in predicting r-process abundances. It is important, when using a model, to confront it to the present observation of shell reduction and re-ordering in regions of medium-mass nuclei, where experimental data such as 2$^+$ energies, reduced transitions probabilities B(E2), atomic masses and single-particle energies exist now.

\section {Proton-neutron forces at the drip-line}  \label{three}
When stellar processes involve nuclei at or close to the drip lines, as it does in neutron-star mergers or in the neutron stars' crust, proton-neutron interactions occur between deeply bound protons and unbound neutrons. The wave functions of the latter explore the continuum and have a poorer radial overlap with the protons confined in the nucleus interior. This probably leads to a globally weaker proton-neutron force than for stable nuclei, in which proton and neutron binding energies are similar. This effect is not yet treated in models and no, or very few, experimental case existed so far to quantify its amplitude as a function of  the nucleon binding energy, angular momentum and coupling to other states in the continuum.  

\begin{figure} [h]
\begin{center}
\includegraphics[width=12cm] {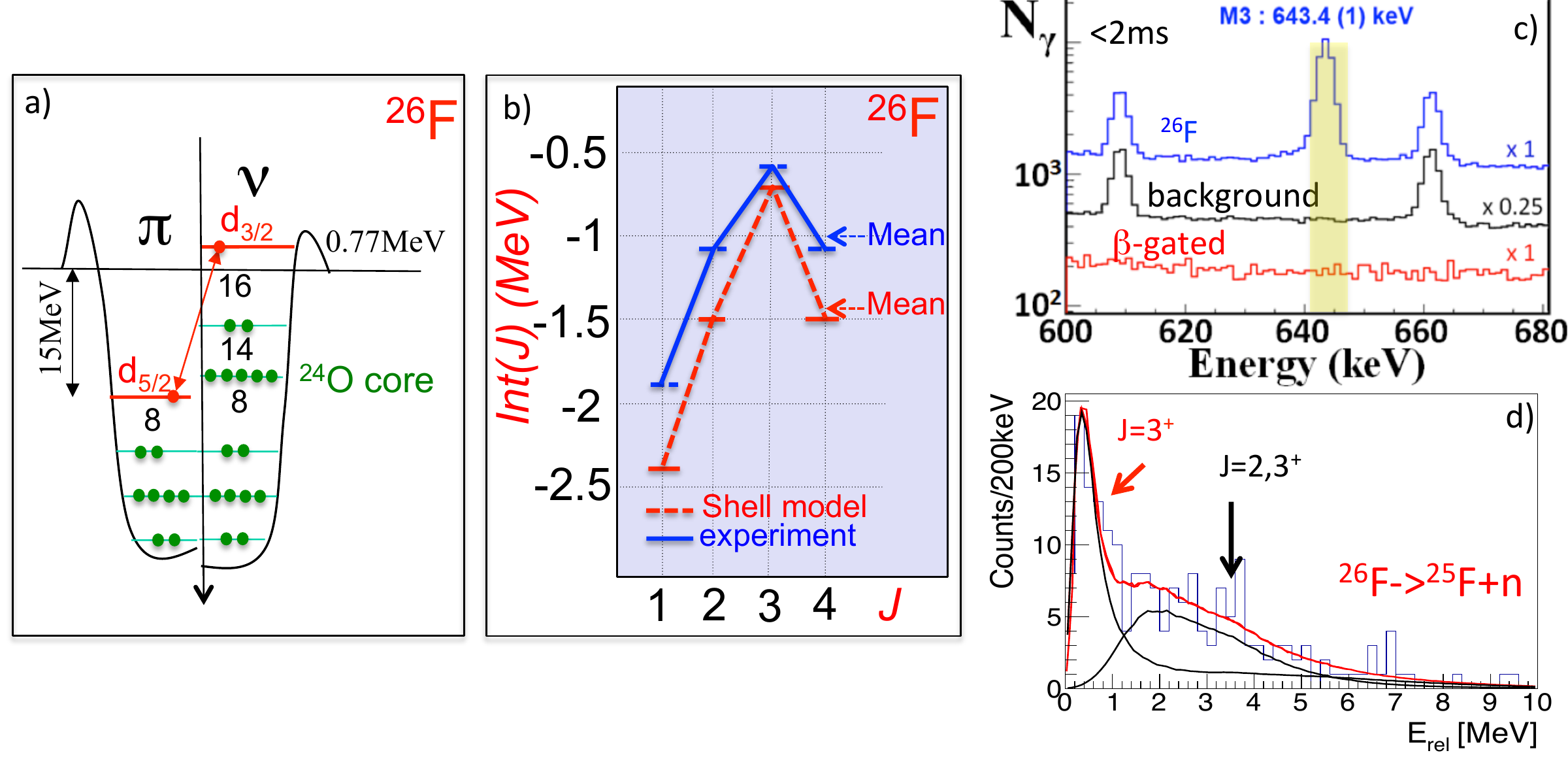}
\caption { a) Schematic picture of the structure of  $^{26}$F nucleus, that can be viewed as a core of  $^{24}$O on top of which a deeply bound proton $d_{5/2}$ and an unbound neutron $d_{3/2}$ interact. Their coupling give rise to an upward parabola of interaction energies $Int(J)$ as a function of $J$ ranging from $1^+$ to 4$^+$, whose calculated (red) and experimental (blue) values, are reported in b). Deviation in mean interaction energy, normalized to $Int(J)$ =0 in case of no interaction, and in the amplitude of the parabola between experiment and theory point to a weaker effective proton neutron interaction at the drip line. c) Experimental $\gamma$ spectrum showing on the top row, in the shaded area, the line corresponding to the decay of 4$^+$ isomer at 643 keV. d) Neutron spectrum $S_n$  showing a resonance at 323(33) keV corresponding to the decay of the 3$^+$ unbound state.
} \label{pn}
\end{center}
\end{figure}

\noindent The $^{26}$F nucleus is one of the few ideal probes to quantify the effect of the proximity of continuum on the effective nuclear force. Indeed, as shown in Fig. \ref{pn}a), it can be viewed as a deeply bound  $d_{5/2}$ proton to which an unbound neutron $d_{3/2}$ of 770 keV\cite{Hoff}  is  added on top of the doubly closed-shell core of $^{24}$O, to form a multiplet of states  $J=1^+ - 4^+$. The binding energies of these four states were required to determine this mean proton-neutron interaction strength, derived from their $J$ weighted-averaged energy values. In Fig.\ref{pn}b), these binding energies are normalized to provide interaction values, $Int(J)$, which are zero in the case that the protons and neutrons do not interact outside the core, and are negative otherwise. Shell model calculations using USDA interaction are shown in the red color curve in Fig.\ref{pn}b). They display an upward parabola as a function of $J$,  whose amplitude and mean interaction amount to about 1.6 MeV and -1.5 MeV, respectively.


\noindent  The energies of the four states of the multiplet were determined using four different  experiments. The ground state binding energy was deduced from its atomic mass value \cite{jura}. The excitation energy of the $J$=2$^+$, 0.657(7) MeV, was obtained using the in-beam $\gamma$-ray spectroscopy technique \cite{Stanoiu}, while a $J$=4$^+$  2 ms-lifetime  isomeric state  was discovered at 0.643 MeV from the observation of its delayed M3 transition to the 1$^+$ ground state \cite{Lepailleur} (Fig. \ref{pn}c). The unbound $J$=3$^+$ state  was populated selectively at the GSI facility by the one-proton knockout reaction from a beam of $^{27}$Ne interacting in a CH$_2$ target placed at the center of the Cristal Ball gamma array \cite{Vandebrouck}. It decayed in-flight by the emission of a neutron ($^{26}$F (3$^+$) $ \rightarrow ^{25}$F+n ), that was detected at forward angles in the segmented neutron array LAND. The momentum vector of the residue $^{25}$F, determined  at the dispersive focal plane of the Aladin spectrometer, was implemented in the invariant mass equation to determine its energy of 323(33) keV above the neutron emission threshold $S_n$ of 1.04(12) MeV (see Fig. \ref{pn}d). The experimental results  reported in the blue curve  of Fig. \ref{pn}b) suggest a weaker mean interaction energy, as well as a more compressed parabola (but still not totally flat), as compared to the USDA shell model calculation \cite{SM} that does not take the proximity of the continuum into account. This comparison points to an effective reduction of the proton-neutron interaction at the drip line by about 30\%. As strongly influencing the shell evolutions and lifetimes values at the drip line, this effect, not yet taken into account in theoretical models, must be more systematically studied and implemented in models to reach satisfactory predictions on the location of the drip lines and the structure of nuclei there. It is noted here that the atomic masses of the nuclei involved to conclude on a reduced strength of the proton neutron force, $^{24}$O, $^{25}$F and $^{26}$F, must be confirmed. 

{\scriptsize
{\bf Acknowledgments} It is a pleasure to address special thanks to the physicists deeply involved in the data analysis of the experiments presented here, such as G. Burgunder, A. Lepailleur, A. Mutchler, V. Vandebrouck, A. Lemasson, K. Wimmer, T. Aumann and C. Caesar. I also warmly thank the whole LISE team at GANIL, the S800/Gretina collaboration at NSCL/MSU, as  well as the ALADIN/LAND collaboration at GSI for their enthusiasm, efficiency and constant support during all the stages of the experiments.}

\end{document}